\documentstyle[aps,prd,epsf]{revtex}
\def\la{\langle} 
\def\ra{\rangle} 
\def\be{\begin{eqnarray}} 
\def\ee{\end{eqnarray}}
\def\zb{\bar{z}}

\newcommand{\eq}{\begin{equation}} \newcommand{\eqx}{\end{equation}}
\newcommand{\dl}{\delta}
\newcommand{\eqn}{\begin{eqnarray}} \newcommand{\eqnx}{\end{eqnarray}}
\newcommand{\f}[2]{\frac{#1}{#2}}

\newcommand{\Tr}{\mbox{\rm Tr}}

\newcommand{\Sg}{\Sigma}
\newcommand{\cor}[1]{\left\langle{#1}\right\rangle}

\renewcommand{\th}{\theta}
\newcommand{\sg}{\sigma}
\newcommand{\lm}{\lambda}
\newcommand{\chis}{\chi_*}
\newcommand{\arr}[4]{
\left(\begin{array}{cc}
#1&#2\\
#3&#4
\end{array}\right)
}
\newcommand{\nn}{\mbox{\bf{}n}}

\begin{document}

\title{$\theta$ Vacuum: A Matrix Model }

\author{Romuald A. Janik$^{1,2}$, Maciej A.  Nowak$^{2,3}$, 
	G\'{a}bor Papp$^{4}$ and Ismail Zahed$^5$}
\address{$^1$ Service de Physique Th\'{e}orique, CEA Saclay, 
F-91191 Gif-Sur-Yvette, France.\\
$^2$ Department of Physics, Jagellonian University, 30-059
Krakow, Poland.\\ $^3$ GSI, Planckstr. 1, D-64291 Darmstadt, Germany\\ 
$^4$CNR Department of Physics, KSU, Kent, Ohio 44242, USA \& \\
Institute for Theoretical Physics, E\"{o}tv\"{o}s University, 
Budapest, Hungary\\
$^5$Department of Physics and Astronomy,
 SUNY, Stony Brook, New York 11794, USA.}
\maketitle

\begin{abstract}
We model the effects of a large number
of zero modes for $N_f$ species of quarks at finite vacuum angle 
$\theta$, using a matrix model with gaussian weights 
constrained by the topological susceptibility and compressibility.
The quenched free energy exhibits a cusp at $\theta <\pi$ 
that is sensitive to the accuracy of the numerical analysis and the 
maximum density of winding modes. Our results bear much in
common with recent lattice simulations by Schierholtz and others. 
The unquenched free energy exhibits similar sensitivities, but for 
small quark masses or a large density of zero modes the results are 
in agreement with those derived using anomalous Ward identitities
and effective Lagrangians.
\end{abstract}

\section{Introduction}

QCD with a finite vacuum angle $\theta$ is subtle.
Canonical quantization~\cite{CANO,COVA} and variational calculations
\cite{VAR} suggest that the vacuum state depends on $\theta$,
while covariant quantization seems to indicate otherwise 
\cite{COVA}. The issue of the $\theta$ angle in QCD and the U(1)
problem are intertwined~\cite{COVA,WARD,WARDC}. At finite $\theta$, 
QCD breaks  CP. Due to the U(1) anomaly, the $\theta$ 
term may be traded from the gauge fields to the quark mass 
matrix. Bounds from the neutron electric-dipole moment yield
$\theta \leq 10^{-9}$ \cite{BALU}.

In QCD the dependence of the vacuum partition function on
the $\theta$ angle involves an understanding of the vacuum 
physics which is essentially nonperturbative. First principle 
calculations are limited and difficult. The reason is that most
lattice QCD simulations rely on important samplings by Monte 
Carlo techniques which require positivity of the action 
configuration by configuration. At finite $\theta$ the action 
is complex in Euclidean space.

Lattice simulations using $CP^n$ models as well as Yang-Mills
theories~\cite{SCHIE,SAM,JAPAN} have been recently carried out, 
with somehow opposite conclusions. A number of effective models 
have been used to gain insights to this important problem
\cite{EFF,WITTEN,OLDTWOFLA}, including recent conjectures
~\cite{WITTENFRESH}. Unfortunately the conventional lore of
power counting, such as chiral perturbation theory, becomes
subtle at finite $\theta$ \cite{COVA,LEUTWYLER}, although some exact
constraints can still be inferred from Ward identities \cite{COVA,
WARD,WARDC}.

In this paper, we will assume that the vacuum supports a $\th$ 
angle and proceed to analyze some related issues using a matrix 
model. The main thrust of our investigation is to try to find out
the conditions under which a numerical analysis of this problem
(albeit in a model) compares to analytical or quasi-analytical
solutions. In sections 2 and 3 we introduce the model, and discuss 
the quenched case. The free energy is found to depend sensitively
on the maximum density of winding modes~\footnote{In the unquenched 
case the number of winding modes is the number of zero modes.} 
${\bf n}$ and the 
accuracy of the numerical calculations. Our observations are similar 
(although not identical) to those reached recently by Schierholtz and 
others~\cite{SCHIE,SAM,JAPAN} using lattice simulations. 
In section~4, we discuss the unquenched case and show that 
under general conditions the saddle-point results agree with the 
numerical calculations. Our conclusions are in section 5. 
Some technical details are given in the Appendices.

\section{Matrix Model}

Consider the partition function described by
\be
  Z(\theta, N_f)=\left\la\prod_{j=1}^{N_f}
                 {\rm det}{\left(\begin{array}{cc}
	im_je^{i\theta/N_f}        & W \cr
	W^\dagger & im_je^{-i\theta/N_f}
  \end{array}\right)}\right\ra \,.
\label{partsumf}
\ee
where the averaging is carried using the weight
\be
\sum_{n_{\pm}} \int dW dW^{\dagger} e^{-\frac 12\, n\,\Tr W^\dagger W}
e^{\frac{-{\chi}^2}{2\chi_* V}}
e^{\frac{-\sigma^2}{2\sigma_*V}}\,.
\label{weight}
\ee
Here $W$ is a complex hermitian asymmetric $n_+\times n_-$ matrix,
$n=n_++n_-$, and $\sigma\pm \chi =2n_{\pm} - \la n\ra$. 
The mean number of zero modes
$\la n\ra$ is either fixed from the outside or evaluated using
the gaussian measure (\ref{weight}). 
For simplicity in this paper we used the quenched measure without the
fermion determinant to fix $\la n\ra$.
Throughout, the value of the quark condensate $\Sigma=1$ in the 
chiral limit. It is readily reinstated by dimensional inspection. 

For a recent review on matrix models in QCD we refer to~\cite{US1}
(and references therein). In short, (\ref{partsumf}) with 
gaussian weights is borrowed from the effective instanton
vacuum analysis~\cite{US} where $n_+$ counts the number of right-handed
zero modes , and $n_-$ the number of left-handed zero modes (restricted
to zero dimension). The number of exact topological zero modes is
commensurate with the net winding number carried by the instantons and
antiinstantons.  
By analogy with~\cite{US}, $\chi_*$ and 
$\sigma_*$ will refer to the unquenched topological susceptibility and particle
compressibility, respectively. Higher susceptibilities 
may be enforced by non-gaussian weights~\cite{ARIEL,DIA}, 
provided that the QCD beta function is restricted to its one-loop form. This 
alternative will be discussed elsewhere.

From~\cite{US} we have $\sigma_*^2=4\nn_*/b$ with $b=11N_c/3$
and $\nn_* =\la n\ra/V$ the mean density of zero modes. If the
compressibility $\sigma_*$ is assumed small in units of 
$\Sigma =1$, then typically $n\sim \la n\ra$. We note that for the
canonical choice $\nn_*=1$, $\sigma_*=0.6$ for $N_c=3$. Finally, through a 
chiral rotation the $\theta$ angle may be removed from the determinant to 
generate an extra phase $e^{i\chi\theta}$ in the measure (see
also~(\ref{jaco})). 
In this form, the $2\pi$ periodicity in $\theta$ is manifest. We recall that in
the original instanton model, CP is explicitly
upset at finite $\theta$, with the exception of $\theta=\pm\pi$ 
(mod $2\pi$)~\footnote{Under CP
$\theta=\pi$ goes to $-\pi$ which is the same as $\pi$ because of the
$2\pi$ periodicity. Mutiple degeneracy of states can however take 
place~\cite{WARDC,WITTEN,OLDTWOFLA}.}.

\section{$N_f=0$}

In this section we will only analyze the quenched partition function
with $N_f=0$, thereby probing the nature of the measure (\ref{weight}).
As we will show, this is not a trivial exercise and the outcome bears
much in common with current quenched lattice simulations. First, we
discuss the case where $\la n\ra=\infty$ with no restriction
on the value of $n$, and hence no restriction on the value of 
$\chi$. Second, we discuss the case where $\la n\ra$ is large but finite, 
so that $|\chi | \leq n$ with typically $n\sim \la n\ra$ for a peaked
distribution in $n$.

\subsection{Infinite Sum}

When the sum is unrestricted and infinite, we have for the quenched
partition function (up to an irrelevant normalization)
\be
Z_Q (\theta) = 	Z (\theta , 0) = \sum_{\chi=-\infty}^{\infty}
e^{i\chi\theta} e^{-{\chi}^2/2V\chi_*} \,.
\label{partq}
\ee 
Using Poisson resummation formula we have
\be
Z_Q(\theta ) = \sum_{k=-\infty}^{+\infty}
e^{-\frac 12 V \chi_* ({\theta} - 2\pi k)^2}
=\theta_3 (\theta/2, e^{-\tau}) \,.
\label{POISSON}
\ee
where the last equality involves the third elliptic $\theta$-function 
with $\tau=1/(2V\chi_*)$. The result is manifestly $2\pi$ periodic. The
vacuum free energy, $F_Q(\theta) =-{\rm ln}Z_Q (\theta)/V$ as 
$V\rightarrow\infty$ is simply
\be
F_Q(\theta ) = {\rm min} \frac 12 \chi_* \left(\theta  +{\rm mod}\, 2\pi
\right)^2
\label{SCISAW}
\ee
in agreement with the saddle-point approximation to (\ref{partq}).
This simple result is the same as the one obtained
using large $N_c$ arguments~\cite{WITTEN}, and  recent
duality arguments~\cite{WITTENFRESH}. 
We observe that the cusp at $\theta=\pi$ (mod $2\pi$)
sets in for $V=\infty$. For finite $V$ the
sums converge uniformly, so that
\be
F_Q' (\theta) =  -\frac 2{VZ_Q(\theta)} 
\sum_{\chi=1}^\infty\ \chi \sin{(\chi\theta)} e^{-\chi^2/2V\chi_*}.
\ee
which is always zero at $\theta =\pi$. As $V\rightarrow\infty$, the 
interchange of the derivative with the sum is not valid, hence the 
cusps. When translated to the quenched instanton calculations, these
cusps indicate a new phase with spontaneous CP violation.
A similar observation was made in the 
context of 1+1 compact electromagnetism using the character 
expansion~\cite{HASSAN}.

Finally, we note that (\ref{POISSON}) can also be rewritten as 
\be
Z_Q(\theta)=Z_E(\theta) +Z_O(\theta) =
\f12 \sqrt{\f{\pi}\tau} \left( \sum_{k=-\infty}^{+\infty}
e^{-\frac1{4\tau} ({\theta} -k\pi)^2} 
+\sum_{k=-\infty}^{+\infty}
(-1)^k e^{-\frac1{4\tau} ( {\theta} -k\pi)^2} \right)
={\cal{\theta}}_3 (\theta , e^{-4\tau}) + 
{\cal{\theta}}_2 (\theta , e^{-4\tau})
\label{ODD}
\ee
where $Z_{E}$ is  $\pi$ periodic and $Z_{O}$ is $\pi$ antiperiodic.
The appearance of $(-1)^k$ in the second sum in (\ref{ODD})
is important for restoring the $2\pi$ periodicity in the full sum.
In the thermodynamical limit, the even and odd
sums are dominated by single Gaussians with 
$F_E=\chi_*\,(\theta -k\pi)^2/2$ for $|\theta/\pi -k|\leq 1/2$,
and $F_O=F_E + ik\pi/V$.

\subsection{Finite Sum}

In the model we are considering the sum over $\chi$ is
restricted to
$|\chi |< N$, with $N={\rm max}\,\, n$. We will denote by 
$\nn =N/V$ the maximum density of winding modes. While in general
$\nn\neq \nn_*$, for a peaked distribution in $n$ (small 
compressibility $\sigma_*$) we expect $\nn\sim \nn_*$. This 
will be assumed throughout unless indicated otherwise. Hence

\be
Z_Q (\theta) = \sum_{\chi=-(N-1)}^{N-1} e^{i\theta\chi} e^{-\chi^2/2V\chi_*}
\label{que-num}
\ee
Approximating the sum in (\ref{que-num}) by an 
integral and evaluating it by saddle point we obtain  
$Z_Q \sim e^{-V\chis\th^2/2}$.

\begin{figure}[htbp]
\centerline{\epsfysize=45mm \epsfbox{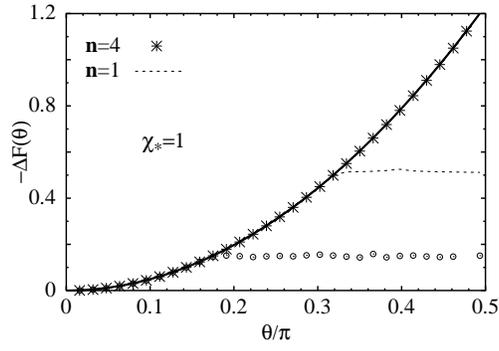}}
\caption{Quenched free energy $\Delta F_Q (\theta )$. See text.}
\label{fig-que}
\end{figure}

In Fig.~\ref{fig-que} we show the numerically generated result versus 
the saddle point approximation for the full free energy
($2\pi$-periodic) and different values of ${\bf n}$.
The normalization was chosen so that $F_Q (0)
=0$. For $N=250$ and $\chi_*=1$, the double precision (16 digit)
numerics (circles) breaks away from the saddle point approximation
(solid line) at $\theta/\pi\sim 0.2$ for both $\nn=1$ and $\nn=4$, while the
high-precision~\cite{HIPREC} (64 digits) calculations (dashed line)
break away at $\theta/\pi\sim 0.3$ at $\nn=1$ but agree with the saddle
point result at $\nn=4$.


The leveling for small values of $\nn$ persists even at infinite
accuracy, and in our case is caused by the  finite range of the
summation over $\chi$ in (\ref{que-num}). Indeed, using the
Euler-MacLaurin summation formula, we have

\eqn
\sum_{\chi=0}^{N-1}f(\chi) - \int_0^{N}f(\chi)d\chi=
-\f{1}{2}[f(0)+f(N)]+
\f{1}{12}[f'(N)-f'(0)]-\f{1}{720}[f'''(N)-f'''(0)]+\ldots
\label{maclau}
\eqnx
where $\ldots$ stand for odd derivatives of $f(N)=2\cos(N\th)
e^{-N^2/2V\chi_*}$. The generic behavior of the correction terms
is a non-exponential prefactor times $e^{-{N^2}/{2\chis V}}$
with no dependence on $\theta$~\footnote{$\theta$
enters only in the prefactor which drops when taking the 
logarithm and dividing out by $V$.}. This should be compared to the 
leading order result, $e^{-V\chis\theta^2/2}$, hence, a breakdown of the 
saddle-point approximation for $N=V$ is expected at 
$\chi_*\theta^2 \sim 1/\chi_*$, that is 
$\theta/\pi\sim 1/\chi_*\pi\approx 0.3$, in agreement with the 
high-precision calculations of Fig.~\ref{fig-que}. 

If we were to increase the density of winding modes to
$\nn\ge\pi\chis$, then for say $\nn =4$, the 
saddle-point result is recovered as indicated by the stars 
(Fig.~\ref{fig-que}). The breaking
point noticed above, now lies outside the period of the free energy (at
$\theta \sim \nn/\chi_*$). Again,
the precision in the numerical calculation is important. For $\nn=4$ the 
precision is upgraded from 100 digits for $\theta/\pi <0.3$ to 150 digits 
for $\theta/\pi > 0.3$. We have observed that for low precision measurements 
(16 digits), the numerical calculations deviate from the saddle-point 
approximation for small values of $\theta/\pi \sim 0.2$ (circles) even for 
$\nn=4$.

\subsection{Comparison to Lattice Simulations}

The observations of the preceding paragraph may be summarized as
follows: the leveling of the quenched free energy of the 
matrix model as a 
function of $\th$ is sensitive to the numerical accuracy of the
calculation. The leveling stabilizes at large numerical accuracy,
and is found to depend on the maximum density of winding modes $\nn$
(typically $\nn\sim \nn_*$) and its relative magnitude to the
topological susceptibility. For the
measure (\ref{weight}), the  
leveling occurs at $\theta_*\sim \nn/\chi_*$ in agreement with 
analytical estimates.

\begin{figure}[tbp]
\centerline{\epsfysize=60mm \epsfbox{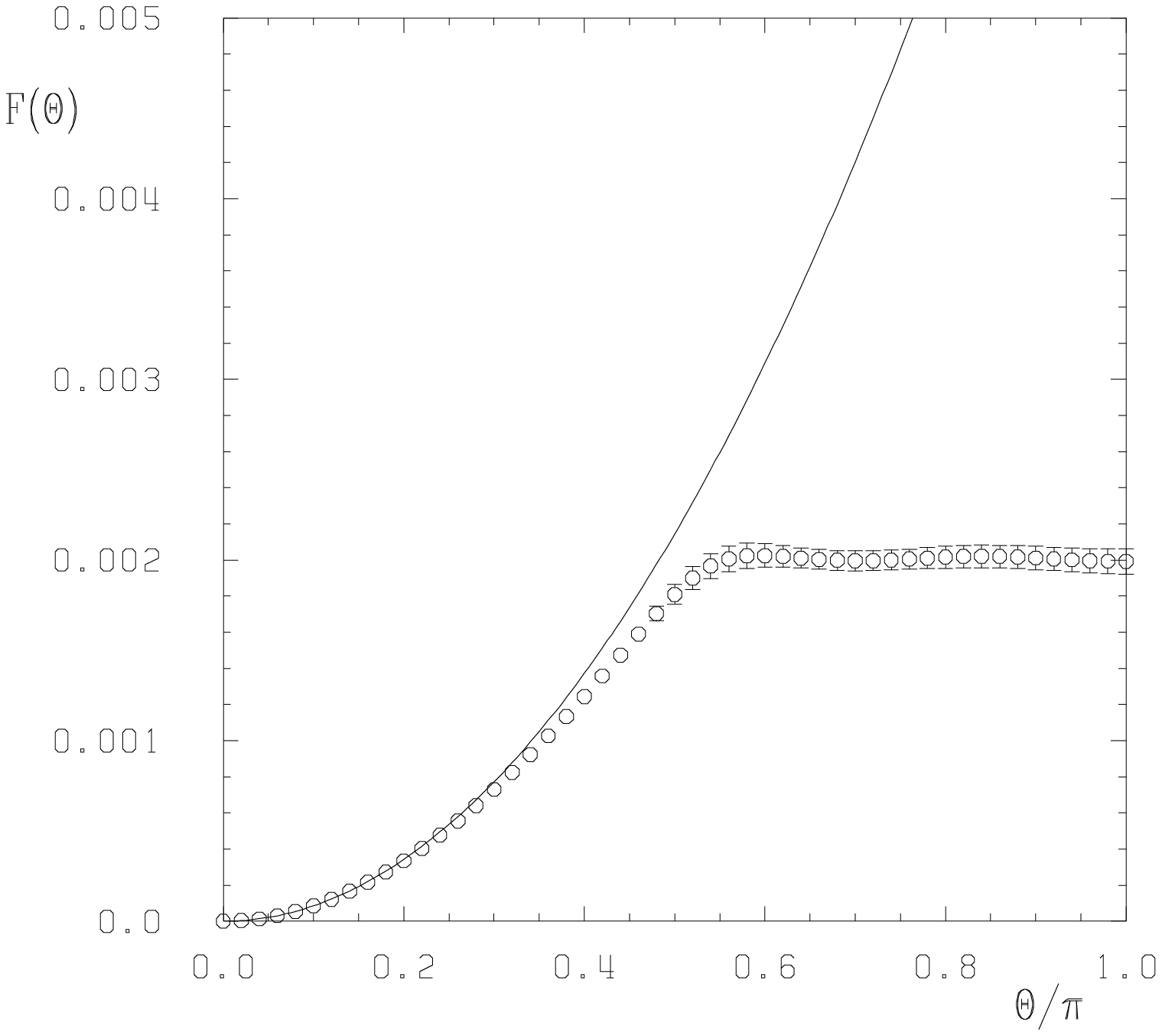} \hspace*{10mm}
 \epsfysize=60mm \epsfbox{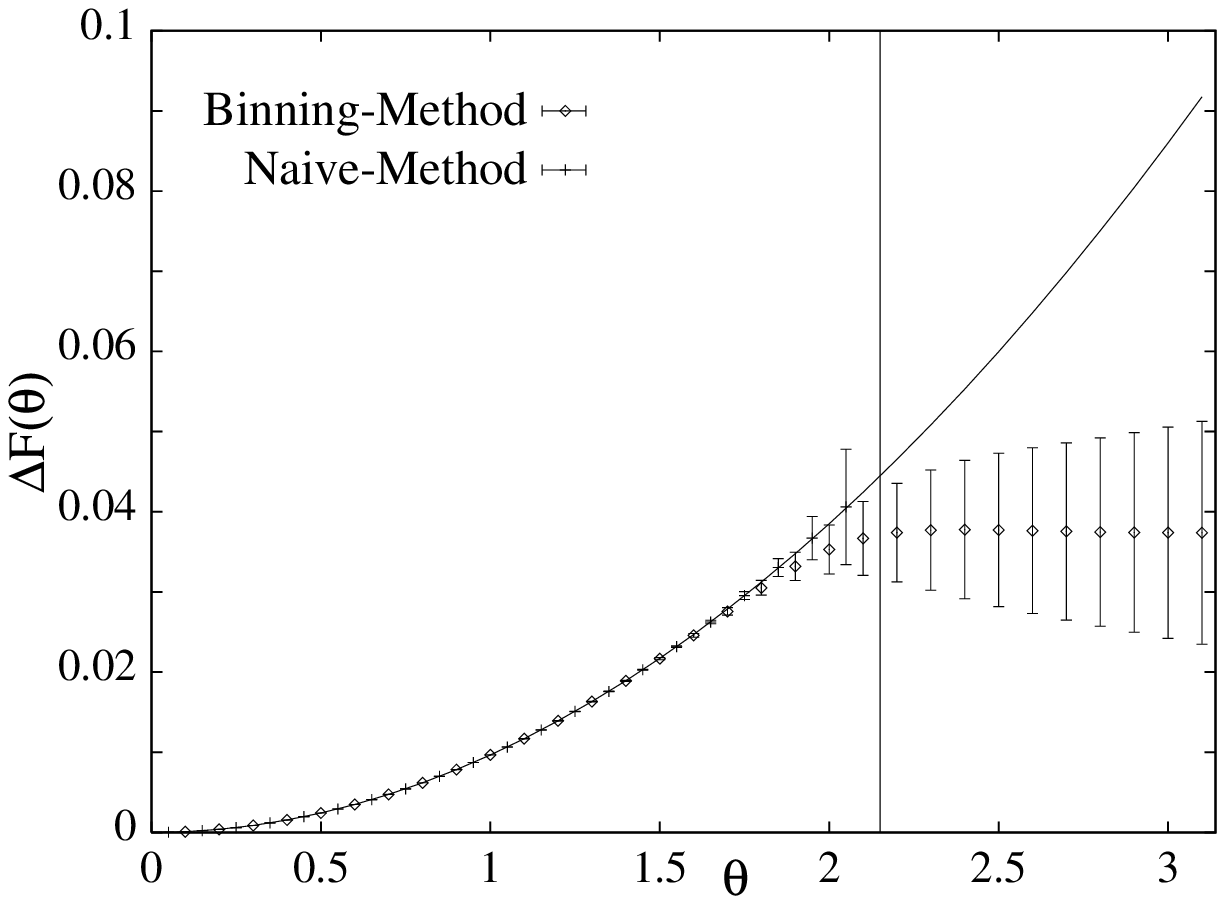}}
\caption{Free energy from a $CP^n$ model 
(left)~\protect\cite{SCHIE} and (right)~\protect\cite{SAM}.}
\label{fig-cpn}
\end{figure}

In an interesting series of investigations, 
Schierholtz~\cite{SCHIE} analyzed numerically
the effects of a finite $\theta$ angle using a $CP^3$ model in two-dimensions,
and also Yang-Mills theory in four-dimensions. The $CP^3$ simulations 
shown in Fig.~\ref{fig-cpn} (left) indicate a leveling of the free-energy 
for $\theta\leq \pi/2$, cautiously interpreted as a possible evidence for a 
first order transition to a CP-symmetric state~\cite{SCHIE}. We note the 
striking similarity of this Figure with Fig.~\ref{fig-que}. 

These findings were recently reexamined by Plefka and Samuel~\cite{SAM} who 
concluded that the apparent first order transition was a possible artifact of
the accuracy of the numerical simulation for fixed lattice size, as shown 
in Fig.~\ref{fig-cpn} (right). Our findings in the matrix model confirm 
this observation. The larger the 
volume $V$, the more precision is needed (exponential precision for infinite 
$V$), due to large cancellations in the partition function caused by the 
oscillating phase $e^{i\chi \theta}$. However, we have also found that the
leveling depends quantitativaly on the maximum density of winding modes
$\nn$ and persists whatever the precision for $\nn\sim 1$. In this sense, 
it would be very useful to understand the dependence of the results in 
\cite{SCHIE,SAM} on $\nn$, with in particular the ones shown in 
Fig.~\ref{fig-cpn}.

\section{$N_f >0$}

To assess the effects of light quarks on the partition function of the 
matrix model, we will consider in this section the general case with
$N_f>0$. We will evaluate the free energy of the matrix model in the
saddle-point approximation after bozonization, and compare the outcome 
to direct numerical calculations using large ensemble of asymmetric
matrices or quasi-analytical methods.

\subsection{Bosonization}

In (\ref{partsumf}) the fermion determinant 
can be rewritten as an integral over $N_fn$-component Grassmanians 
$\psi = (\psi_R, \psi_L)$ with dimensions $(n_+, n_-)$, that is

\be
{\rm det}_F = \int d\psi d\psi^{\dagger}
e^{\psi^{\dagger}_R m e^{i\theta/N_f} \psi_R + \psi^{\dagger}_R iW\psi_L + 
L\leftrightarrow R} \,.
\label{boso}
\ee
The $\theta$ angle can then be removed from the action by a U(1) 
transformation $\psi\rightarrow e^{i\gamma_5\theta/2N_F} q$ with 
$\gamma_5 = {\rm diag} ({\bf 1}_{n_+}, -{\bf 1}_{n_-})$. The 
Jacobian of this transformation is just 

\be
{\rm det}_F\,e^{i\gamma_5\theta/N_f} = e^{i\theta (n_+-n_-)}
\label{jaco}
\ee
which is the well known trade-off through the U(1) anomaly. As a 
result, we have the following Ward identity
\be
\left\la n_+-n_-\right\ra_{\theta}  = m \left\la 
q^{\dagger}i\gamma_5 q\right\ra_{\theta}\,.
\label{ward}
\ee
Assuming that the range of resummation over $\chi=n_+-n_-$ is 
infinite, that the distribution in $n=n_++n_-$ is peaked ($n\rightarrow
\la n\ra$) 
and trading the $\chi$-sum by an integral in (\ref{partsumf}), we 
have,

\be
  Z (\theta , N_f) = \int dP dP^\dagger exp\left\{ \la n\ra \left[
	-\frac 12{\rm Tr\,}|P|^2 + \frac 12 {\rm Tr\,} 
	\log{|z+P|^2}
+\frac {\chi_*}{2\nn_*}\left( \log \left({\rm det}{\frac{z+P}%
	{\overline{z} +P^{\dagger}}}\right)^{\frac 12} \right)^2 \right] \right\}
\label{AP1}
\ee
where $z={\rm diag}\,\,\, m_je^{i\theta/N_f}$.
The logarithm in (\ref{AP1}) is multivalued and
its $2\pi$ determination will be assumed. 
In analogy to the quenched case, the approximation of the sum by an integral
should be valid for $\nn \sim\nn_* >\pi\chi_{top}$, where we substituted
the quenched topological susceptibiliy $\chis$, by the unquenched
one $\chi_{top}\le\chis$. For small masses $\chi_{top}$ gets screened
(see Appendix C).

To keep a tab on the validity
of trading the $\chi$-sum by an integral, we observe that for
$N_f=1$ and $z=me^{i\theta}$, an exact form can be reached for
\be
Z (\theta, 1 ) = 
\left\la (z + P)^{n_+} (\zb + P^{\dagger})^{n_-}\right\ra
\label{part}
\ee
followed by the substitution $P\rightarrow W$ in the measure (\ref{weight}).
Here $P$ is a complex variable. Because of (\ref{jaco}), the same partition
function can be written as
\be
Z(\theta , 1) = \sum_{\chi=-\infty}^{\infty}\ e^{i\chi\theta} 
	e^{-\chi^2/2V\chi_*} Z_\chi
\label{zfull}
\ee
where $Z_\chi$
is the partition function for fixed asymmetry $\chi$. Again for
a peaked distribution $N\sim \la n\ra$, so that
\be
Z_{\chi}= \left(\frac2N\right)^{\frac{N-\chi+2}2} \pi m^{\chi} e^{-Nm^2/2}
 \frac{\Gamma\left(\frac{N+\chi+2}2\right)}{\Gamma(\chi+1)}
  {}_1F_1 (\frac{N+\chi+2}2, \chi+1;Nm^2/2) \,.
\label{zchia}
\ee
where ${}_1F_1$ is Kummer's (confluent hypergeometric) function.
For even $\chi$, it reduces to an associated Laguerre polynomial
in $Nm^2$. We note that for $N\rightarrow\infty$ with $Nm$ and $\chi$
fixed, $Z_{\chi}\sim I_{\chi} (Nm)$ which is the expected 
generating function for the microscopic sum rules~\cite{LEUTEFF}.

\subsection{Saddle-Point Approximation}

Without loss of generality, we can set
$P={\rm diag}\,\,\,p_je^{i(\theta/N_f-\phi_j)}$, 
so that the unsubtracted free energy associated 
to (\ref{AP1}) reads
\be
F = \f{\nn_*}{2} \sum_{j=1}^{N_f} \Big[ p_j^2
-\log{(p_j^2 + 2m_j p_j\cos{\phi_j} + m_j^2)}\Big]
	+\frac {\chi_* }{2} \left(\theta   + 
	 {\rm arg}\left( \prod_{j=1}^{N_f}{\left(
	\frac{m_j+p_je^{-i\phi_j}}{m_j+p_je^{i\phi_j}} \right)}\right)^{\frac 
        12} +{\rm mod}\,2\pi \right)^2 \,.
\label{AP2}
\ee
We note that for $N_f=0$ (\ref{AP2}) reduces to~(\ref{SCISAW}). 
For small masses $m_j\ll \chi_*\sim p_j\sim 1$ and
(\ref{AP2}) simplifies to
\be
F = \f{\nn_*}{2} \sum_{j=1}^{N_f} \left( p_j^2 -\log{p_j^2} -2
	\frac{m_j}{p_j} \cos{\phi_j} \right) 
	+\frac 12 \chis \left(\left(\theta 
        -\sum_{j=1}^{N_f} \phi_j +{\rm mod}\,2\pi \right)
	+\sum_{j=1}^{N_f} \frac{m_j}{p_j} \sin{\phi_j} 
        \right)^2
\label{AP3}
\ee
to order ${\cal O}(m^2)$.
The saddle point in the $p$'s decouples and gives
$p_j=1-m_j/2 \cos{\th}+{\cal O}(m^2)$. To the same order,
the saddle point in the $\phi$'s is
\be
\theta=\sum_{j=1}^{N_f} \phi_j +{\cal O} (m)
	\quad\mbox{and}\quad
m_1{\rm sin}\, \phi_1= ...=m_{N_f} {\rm sin}\, \phi_{N_f} \,.
\label{AP4}
\ee
These equations were derived using 
large $N_c$ arguments~\cite{WITTEN,OLDTWOFLA} and 
anomalous Ward-identities~\cite{WARDC}.

For $N_f=1$, we have
\be
\phi=\theta +m
\left(\frac {\nn_*}{\chis} -1\right) \sin{\theta} +{\cal O}(m^2)
\label{APP4}
\ee
for which the subtracted free energy is 
$\Delta F=\nn_* m(1-\cos{\theta})$.

For $N_f=2$, we have
\be
{\rm sin}\phi_{1,2} =\pm \frac {m_{2,1} {\rm sin}\, \theta}
{\sqrt{m_1^2 +m_2^2 + 2m_1m_2 {\rm cos}\,\theta}}
\label{AP5}
\ee
and the subtracted free energy now reads
\be
\f{1}{\nn_*}\Delta F(\theta) = |m_1+m_2| -\sqrt{m_1^2+m_2^2 + 2m_1m_2 
\cos{\theta}} \,.
\label{FREETWO}
\ee
For $m_1=m_2$ a cusp develops at $\theta=\pi$, since $\Delta F(\theta ) =\nn_*|m| 
(1-|\cos{\theta/2} | )$. In this case both numerator and denominator
vanish in (\ref{AP5}), hence any value of $\phi_1$ and $\phi_2$ is allowed 
provided that $\phi_1+\phi_2=\pi$. A similar behavior was noted 
by many~\cite{WARDC,WITTEN,OLDTWOFLA,CREUTZ,SMILGA}, 
following the spontaneous breaking of strong CP.

For $N_f=3$, the explicit solutions to (\ref{AP4}) 
are in general involved, thereby making an analytical form for the free
energy involved. At $\theta=\pi$, however, the analysis simplifies. 
Using (\ref{AP4}) we obtain a trivial solution with one of the $\phi$'s
being $\pi$ and the others zero, and a non-trivial one,
\be
\cos{\phi_3} = \frac{m_3^2 m_1^2+m_3^2 m_2^2-m_1^2 m_2^2}%
	{2 m_3^2 m_1 m_2}
\label{AP6}
\ee
which is doubly degenerate for
\be
m_1m_2 > m_3 |m_1-m_2|\,.
\label{CONDTHREE}
\ee
The trivial solution corresponds to no cusp at $\theta=\pi$, while the non-trivial
one yields a cusp at $\theta=\pi$ because of the double degeneracy. Again
similar observations were made using effective Lagrangians 
\cite{WITTEN,CREUTZ,SMILGA} and anomalous Ward 
identities~\cite{WARDC}, where (\ref{CONDTHREE}) is known as Dashen's
condition~\cite{DASHEN}. For sufficient flavor breaking (\ref{CONDTHREE})
is not fulfilled and strong CP is not spontaneously broken.
This is the case in nature where $m_3\gg m_1, m_2$.

\subsection{Numerical Analysis for $N_f=1$}

\begin{figure}[htbp]
\centerline{\epsfysize=45mm \epsfbox{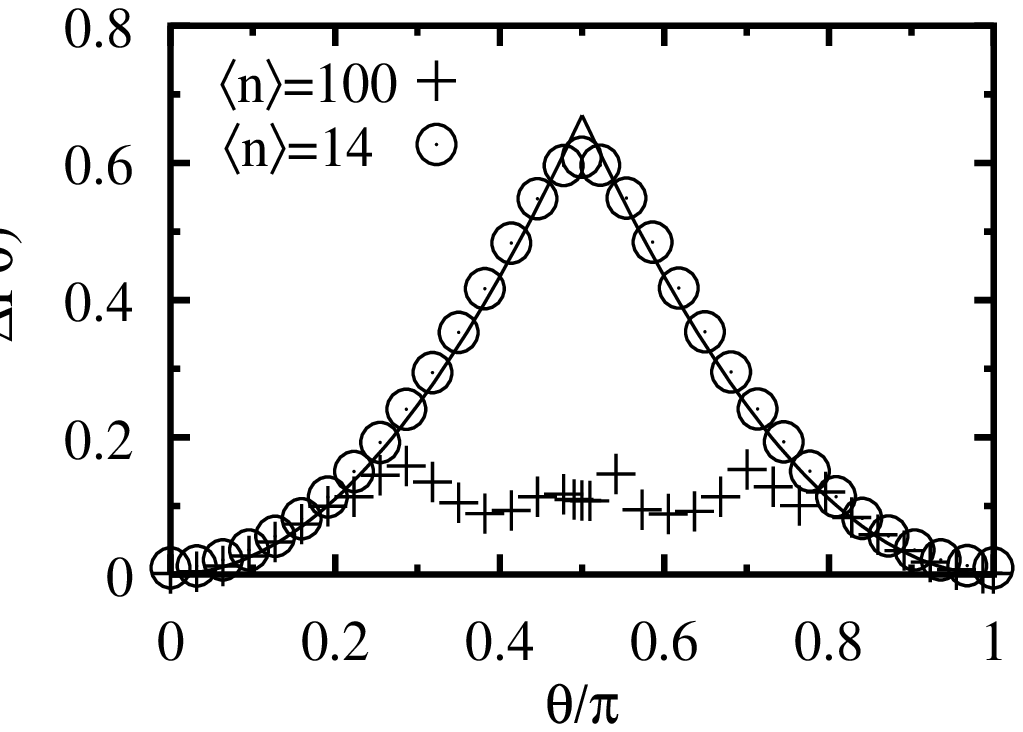} \vspace*{10mm}
\epsfysize=45mm \epsfbox{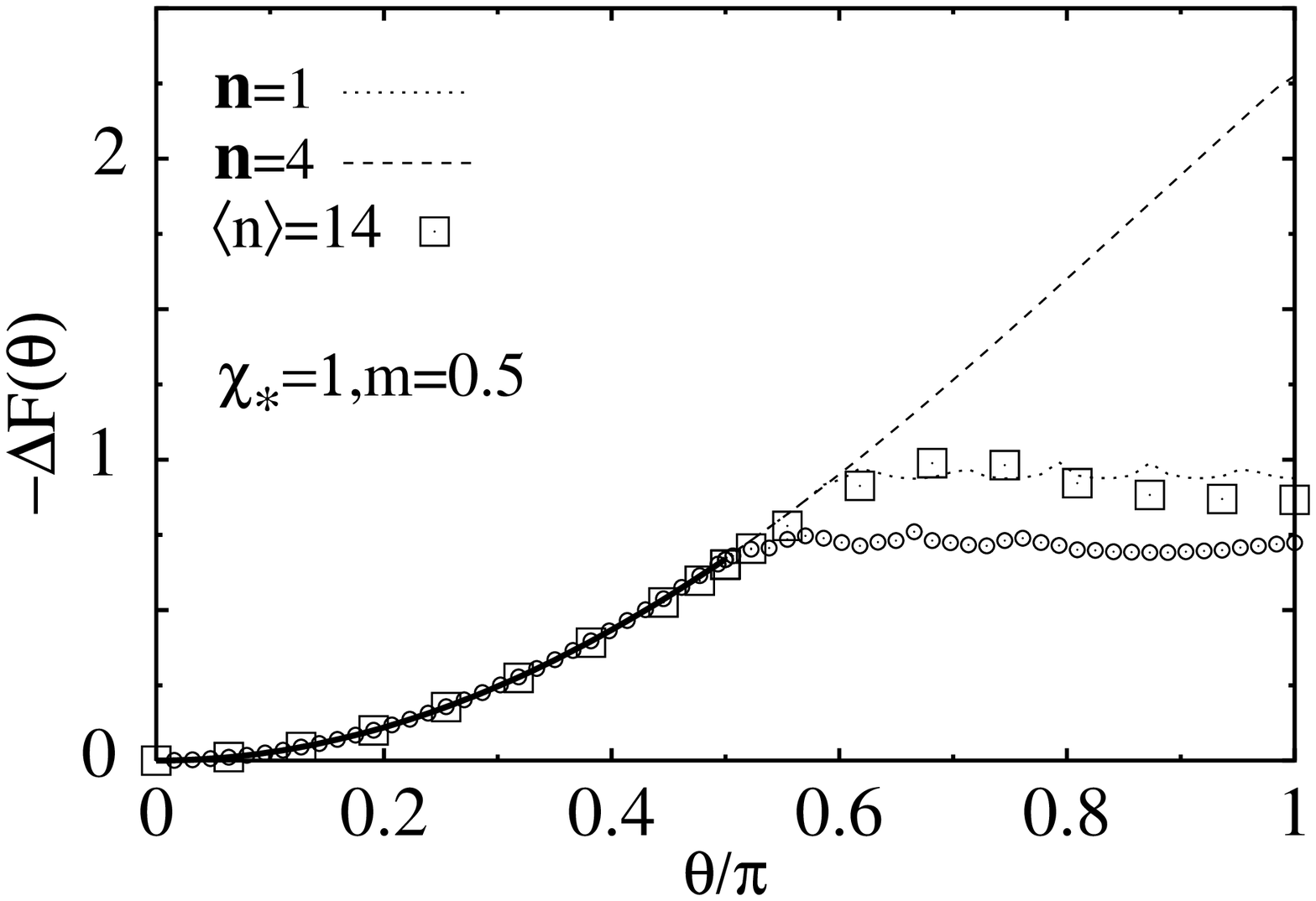}}
\caption{Unquenched free energy for fixed size matrices (left) and
varying size matrices (right) at intermediate mass.}
\label{fig-unqu}
\end{figure}

In Fig.~\ref{fig-unqu} (left) we compare the fixed size
results for the free energy with  $\la n\ra=14$
(circle) and $\la n\ra=100$ (plus) over one 
period~\footnote{Since $\la n\ra$ is fixed even, the periodicity
is $\pi$ and not $2\pi$ as noted in section 3.1.},
using single precision (16 digits) numerical calculations for 
$m=0.5, \chis=1$. The saddle point solution is indicated by the solid line. 
Increasing the size of the matrices $worsen$ 
the agreement with the saddle-point result (solid line). 
The reason is that for $\pi/2$ the integrand in the  sum develops 
alternating signs with large numerical cancellations. The breakdown 
of the numerical calculation takes place at 
$\theta_{*} \sim -(\ln \,\, \epsilon)/(\langle n\rangle\chi^*)$
with $\epsilon$ the numerical precision. 
These observations are new as they pertain to the unquenched free
energy.

In Fig.~\ref{fig-unqu} (right) we show the same results after averaging
over the size of the matrices (with $\sigma_*$=1),
restoring the $2\pi$ periodicity. 
Changing the gaussian measure in $n$ to a uniform one with $|n-\la n\ra|
\le 6$ does not make a noticeable change on the Figure.
We compare the low (16 digits, circles) and high (64
digits, boxes) precision calculations for an {\em averaged} size $\la
n\ra=14$ using an ensemble of 500000 matrices.
Large size samplings at large $\theta$ requires {\em exponentially} larger
precision. Much like the quenched case we observe the same dependence on 
$\nn\sim\nn_*$, causing a departure from the saddle point results at 
$\theta\sim 0.6\pi$ for $\nn =1$.

Since the $\nn >1$ region is unaccessible by direct use of random
matrices, we will use the partition 
function for fixed asymmetry $Z_{\chi}$ in~(\ref{zfull}) in the saddle point 
approximation and perform the final sum in~(\ref{zfull}) numerically. We have
checked that the saddle-point result agrees with the result obtained by
averaging over large ensembles of matrices with fixed $\chi$ for $\nn=1$. 
We refer to this analysis as quasi-analytical and the results are shown in
Fig.~\ref{fig-unqu} (right) for $\nn=1$ (dotted line) and $\nn=4$ (dashed 
line). 

\begin{figure}[htbp]
\label{DER}
\centerline{\epsfysize=45mm \epsfbox{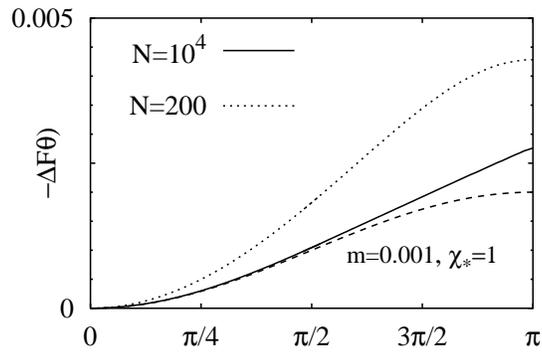}}
\caption{Unquenched free energy $\Delta F (\th )$ for $N_f=1$. See text.}
\label{fig-unq001}
\end{figure}

For small masses and $\th <\pi/2$ the free energy in the saddle-point
approximation is $F(\theta )\sim \nn_* m (1- \cos{\theta} )$ with no
cusp whatever $\theta$.  For large
$m$ a cusp at $\theta=\pi$ develops, following the decoupling of the
flavor (roughly $m\sim N_f\chi_*$). In Fig.~\ref{fig-unqu} the
occurence of a cusp for $m=0.5$ and $\nn =4$ is an artifact of the
quasi-analytical analysis where only the saddle-point in $\chi$ is
retained (see above). Indeed, the numerical results from this procedure 
are compared with the standard saddle-point approximation (see previous 
section) in Fig.~\ref{fig-unq001} for $m=0.001$ (small mass) and 
$\nn_*=\nn=1$ ($\la n\ra=N$). 

We may now ask if a levelling in the free energy occurs
for small masses $m$. As we wrote in section 4.1 this 
takes place for $\nn<\pi\chi_{top}$. However, due to the
screening of the topological charge by the fermion determinant for
small masses we expect $\chi_{top}\sim m$ (see Appendix C) so that
the levelling is ruled out for reasonable values of $\nn$. We note that
the macroscopic limit is reached only for $Nm$ large. For $N=200$ ($Nm<1$) 
many modes are still missing in the sum~(\ref{zfull}) 
(dotted line). For $N=10^{4}$ ($Nm\gg 1$) and $\theta<\pi/2$,
the quasi-analytical procedure (solid) and the saddle-point approximation 
(dashed) are in agreement.

Finally, we now ask whether the standard saddle point method fails
for $\th >\pi/2$, leading possibly to a cusp at $\th\leq\pi$, 
contrary to expectations. For $N_f=1$ we can calculate the fixed
asymmetry partition function exactly~(\ref{zchia}) and use it
for a comparison with the quasi-analytical procedure (see above).
The outcome confirms the standard saddle point result with no
cusp at $\theta=\pi$, and infirms the quasi-analytical
result for large values of $\th$. We conclude that
the saddle-point approximation carried prior to the
$\chi$-resummation is only valid for $\th$ small (the
$1/N$ terms at large $\th$ are important), while the one
carried after the $\chi$-resummation is valid whathever $\th$.


We have numerically checked, that most of the present observations carry 
to $N_f>1$. In particular, a cusp may form in the latter for sufficiently 
degenerate quark masses, in agreement with the saddle point analysis 
discussed above.

\section{Conclusions}

We have analyzed the effects of a finite vacuum angle $\theta$ on the
vacuum partition function described by a matrix model, both in the 
quenched and unquenched approximation. The results are subtle. 

In the quenched case, we have found that the free energy exhibits a cusp 
at finite $\theta$ that is sensitive to the precision of the numerical
analysis. On this point, we are in agreement with the lattice 
analysis~\cite{SAM}. However, we have further noticed that the results
are also sensitive to the maximum density of winding modes $\nn$. For
a small compressibility or a peaked distribution in $n$,
$\nn$ is similar to $\nn_*$, the mean winding
density. For large enough $\nn >1$ the position of the cusp 
is moved to $\theta=\pi$ for high enough precision. This observation 
may be of relevance to the lattice results~\cite{SCHIE,SAM}. In this context,
it would be interesting to compare the lattice distributions for $n_{\pm}$
in~\cite{SCHIE,SAM} to the gaussian ones we have used in our work.

In the unquenched case, a similar dependence on ${\bf n}$ is found,
where $\nn$ is also interpreted as the maximum density of zero modes.
For sufficiently large ${\bf n}$ and large masses the quenched results 
are recovered. Each mass decouples at $m\sim \chi_*$, although for
large and degenerate masses $m\sim N_f\chi_*$ (in units where the quark
condensate is one). For $N_f=1$ and small masses with $mV >1$, the 
screening of the topological charge takes place, and the saddle 
point solution holds without any cusp. For $N_f=2,3$, the numerical results 
are found to agree with a saddle-point analysis, and results from anomalous
Ward identities and effective Lagrangians. A  cusp at $\th=\pi$ occurs for 
sufficiently degenerate quark masses. We have found that the use of the 
saddle-point approximation at large $\th$ requires care.

\acknowledgments

We  would like to thank G. Shierholtz for discussions and R. Crewther
for comments. This work was supported in part  by the US DOE grants DE-FG-88ER40388
and DE-FG02-86ER40251,
by the Polish Government Project (KBN)  grant 
2P03B00814 and by the Hungarian grant OTKA-F026622.

\appendix
\section{Alternative Saddle-Point with $N_f=1$}

An alternative saddle-point analysis can be directly performed
for fixed $\chi$ using representation~(\ref{zfull}). For that, we define
\eq
\chi= -in\cdot y \,.
\eqx
The solution for the saddle point equations gives
\eqn
P_{sp}&=&\f{-m^2\pm\sqrt{m^4+4(m^2-y^2)}}{2\zb}+\f{iy}{\zb}\,, \nonumber \\
\label{psaddle}
P^\dagger_{sp}&=&\f{-m^2\pm\sqrt{m^4+4(m^2-y^2)}}{2z}-\f{iy}{z}\,.
\label{saddlepp}
\eqnx
For a peaked distribution in $n$ we expect $n\sim \la n\ra$.
The $P_{sp}$'s are related to the condensate by
\eq
-i\langle q^\dagger q\rangle = 
  \frac 1V \partial_m\log Z=
	\frac {\nn_*}2 (P_{sp}^\dagger e^{i\theta}+P_{sp} e^{-i\theta})\,.
\eqx
The value of $y$ is fixed by requiring the vanishing of the term
proportional to $\chi$. Hence, the consistency condition reads
\eq
\label{consist}
\log \f{D+iy}{D-iy}+2i\th+2i\f{y\nn_*}{\chis}=0
  \quad\mbox{or}\quad
\arctan\f{y}{D}+\th+\f{y\nn_*}{\chis} = 0
\eqx
where $D=(m^2\pm\sqrt{m^4+4(m^2-y^2)})/2$, with $y$ satisfying
\eq
\cor{n_+-n_-}=\nn_* V \,\frac m2\,
(P_{sp}^\dagger e^{i\theta}-P_{sp} e^{-i\theta}) 
	= \nn_*\, y.
\eqx
In the consistency condition, the principal branch of the logarithm is 
retained, making the saddle-point result manifestly $2\pi$ periodic in
$\theta$. We note that the present derivation is equivalent to performing 
the saddle-point calculation for $y$ without substituting the 
form~(\ref{psaddle}) for $P_{sp}$'s. The associated free energy is
\be
F =- \f{1}{V}\log Z (\theta )
=-\frac {\nn_*}2 
       \!\left( \log\left[
	\f{D^2\!+\!y^2}{m^2}\right]
	\!-\!\f{1\!+\!y^2}{D^2\!+\!y^2}m^2\!-\!\f{\nn_*}{\chis}y^2 
	\right) \,.
\ee
The subtracted free energy is $V\Delta 
F(\theta ) = -{\rm ln}\,\, Z(\theta )/Z( 0)$. 

\section{Topological Density}

The topological density $\la n_+ -n_-\ra$ measures the difference between
the number of zero modes with plus and minus charges, in the volume $V$
fixed by the width of the quenched topological susceptibility $\chi_*$. 
This is also the amount of U(1) charge in the vacuum state thanks to 
(\ref{ward}). At the saddle point ($N_f=1$)
\be
\la n_+ -n_-\ra = i \partial_{\theta}\,{\rm ln}\, Z (\theta )/V =
\nn_* V\,\frac m2\left(P_{sp}^{\dagger} e^{i\theta} -P_{sp} e^{-i\theta}
	\right) =  \nn_*\, y.
\label{AP7}
\ee
We see that $y$ measures directly the topological density at the saddle point. 
For a large mass $m$ 
\be
y=-\frac {\theta}{\frac 1D + \frac{\nn_*}{\chi_*}}\sim 
	-\frac 1{\nn_*}\chi_*\theta \left( 1-\frac{\chi_*}{\nn_* m^2} \right)
\label{AP8}
\ee
while for a small mass $m$
\be
\label{ysmall}
\pm y=-m\tan\th |\cos \th |  -m^2
	\f{\chis-2\nn_*}{4\chis} \sin{2\th} + {\cal O}(m^3)\,.
\label{AP9}
\ee
 The $\pm$ solutions correspond to the transformation $\th
 \leftrightarrow  -\th$, whereas the absolute value corresponds to the 
change of the branch in the solution of the consistency equation.

\section{Topological Susceptibility}

The topological susceptibility measures the variance of $(n_+-n_-)$ in the
vacuum state at finite $\theta$. It is simply $\chi_{\rm top} =-\partial 
y/\partial\theta\ \nn_*$. In the large mass limit
\be
\chi_{top} =
\chi_* \left( 1-\frac{\chi_*}{\nn_* m^2} \right) 
	+ {\cal O}(m^{-3})\,,
\label{AP99}
\ee
giving $\chi_{\rm top} =\chi_*$ in the quenched case ($m=\infty$).
In the small mass limit and $|\theta|<\pi/2$, 
\be
\chi_{\rm top} = \nn_* m \,|\cos\th|
	+\nn_* m^2 \f{\chis-2\nn_*}{2\chis} \cos{2\th} + {\cal O}(m^3)\,.
\label{AP10}
\ee
For $\th >\pi/2$ we have $\cor{(n_+-n_-)^2}<0 $,
which is possible since the measure is not semi-definite.

The quark condensate for large masses is
\be
  -i\la q^{\dagger} q \ra = \f{\nn_*}{m} \left(
	1-\f{\chis^2}{\nn_*^2 m^2} \th^2 \right) \,
\ee
and for small masses is 
\be
-i\la q^{\dagger} q \ra = \nn_*\, {\rm cos} \th 
	-\nn_* m \left( \frac 12+\f{\chis-2\nn_*}{2\chis} \sin^2\th \right)
	+{\cal O}(m^2)
\label{AP11}
\ee

The anomalous U(1) Ward identity is given by
\be
\chi_{top} = -im \la q^{\dagger} q\ra -
m^2 \underbrace{\la q^{\dagger} \gamma_5 q \,\,q^{\dagger} 
	\gamma_5 q \ra}_{\chi_{ps}} \,.
\label{AP12}
\ee
For small masses, the insertion of (\ref{AP10}-\ref{AP11}) into
(\ref{AP12}) yield
\be
\chi_{ps} =  -\nn_* \left( \frac 12+\f{\chis-2\nn_*}{2\chis} \cos^2\th \right)
\label{AP13}
\ee
for the pseudoscalar correlator (rightmost term in (\ref{AP12})).
It is finite in the chiral limit. For $\theta=0$ 
we recover the result~\cite{U1US}. 
For large enough $\nn_*>\chis$ at
certain finite angles the pseudoscalar correlator becomes zero.
For large masses we have $\chi_{ps}=(\nn_*-\chis)/m^2$.
The present relations generalize readily to $N_f>1$.

\section{Resolvent for Fixed $\chi$}

The resolvent for the matrix model considered here reads
\eq
G(z)=\cor{\f{1}{2N}\Tr\f{1}{z-\arr{ime^{i\th}}{W}{W^\dagger}{ime^{-i\th}}}}
\eqx
Since the overlap matrix elements do not mix different flavors the
resolvent splits into a sum of 1-flavor resolvents and
we get effectively $N_f$ copies of the appropriate 1-flavor
eigenvalue distributions. 

Since the matrix is nonhermitian it turns out that the eigenvalues
lie on a curve (more precisely on two intervals - see below). This
comes from the decomposition ($N_f=1$)
\eq
\arr{ime^{i\th}}{W}{W^\dagger}{ime^{-i\th}}=
im\cos\th\,\,{\bf 1}+\arr{-m \sin\th}{W}{W^\dagger}{m\sin\th}
\eqx
So the eigenvalues are just the eigenvalues of the two-level hermitian
chiral system displaced by $im\cos\th$. We will write $z'=z-im\cos\th$
and introduce the self energies defined by
\eq
\cor{\f{1}{z-\arr{ime^{i\th}}{W}{W^\dagger}{ime^{-i\th}}}}
=\arr{\f{1}{z-\Sg_1}}{0}{0}{\f{1}{z-\Sg_2}}
\eqx
Taking into account the fact that the random matrices are asymmetric,
we obtain the following equation for the self-energies
\eqn
\Sg_1&=& \f{1-\f{x}{2}}{z'-m\sin\th-\Sg_2}\\
\Sg_2&=& \f{1+\f{x}{2}}{z'+m\sin\th-\Sg_1}
\eqnx
with $x=\chi/N$.
The trace of the resolvent is just $(\Sg_1+\Sg_2)/2$. Hence
\eq
G(z)=z'\cdot
\f{1-\sqrt{1-\f{4}{\sg}+\f{x^2}{\sg^2}}}{2}-\f{m\sin\th}{2\sg}x
\eqx
where $\sg=z'^2-m^2\sin^2\th$.
The eigenvalues lie on two intervals determined by
\eq
z'^2_{cut}-m^2\sin^2\th=\f{x^2}{2\pm\sqrt{4-x^2}}
\eqx
and, for asymmetric matrices, there are additional Dirac delta spikes
at $z=m\sin\th$ when $\chi$ is negative and at $z=-m\sin\th$ when
$\chi$ is positive.
Explicitly the eigenvalue distribution reads
($\lambda =\lambda_R+i\lambda_I$)
\eqn
\nu(\lm) =\dl(\lm_I-m\cos\th)\biggl\{\chi\dl(\lm_R+m\sin\th)+\f{1}{2\pi}
(\lm_R-m\sin\th)\sqrt{ 1-\f{4}{\sg}+\f{x^2}{\sg^2}}\biggr\}
\eqnx
for $\chi$ positive and 
\eqn
\nu(\lm) =\dl(\lm_I-m\cos\th)\biggl\{|\chi|\dl(\lm_R-m\sin\th)
+\f{1}{2\pi}
(\lm_R-m\sin\th)\sqrt{ 1-\f{4}{\sg}+\f{x^2}{\sg^2}}\biggr\}
\eqnx
for $\chi$ negative. The structure of these distributions is not universal,
but may be useful for understanding the $\theta$ structure from the bulk
QCD spectrum using cooled lattice gauge configurations.

\setlength{\baselineskip}{15pt}

\newpage

\vfill

\vfill

\vfill

\vfill
\end{document}